\documentclass[12pt]{article}
\usepackage{rotating,amssymb}
\begin{document}

\thispagestyle{empty}
\hspace*{\fill}  \mbox{WUE-ITP-97-042}\\
\hspace*{\fill}  \mbox{hep-ph/9710519}\\
\vspace*{2 cm}
\begin{center}
{\Large\bfseries Production of neutralinos in superstring-inspired \\
$\mathbf{E_6}$ models\footnote{
Supported by the Deutsche Forschungsgemeinschaft under
contract no.\ FR~1064/2-1 and by the German Federal Ministry for
Research and Technology (BMBF) under contract no.\ 05~7WZ91P~(0).
Contribution to the proceedings of the ``ECFA/DESY Study on Physics and
Detectors for the Linear Collider'', DESY~97-123E, ed.\ R.~Settles}} 
\\[3ex]
{\large S.~Hesselbach\footnote{email:
hesselb@physik.uni-wuerzburg.de}, F.~Franke\footnote{email:
fabian@physik.uni-wuerzburg.de} , H.~Fraas\footnote{email:
fraas@physik.uni-wuerzburg.de} 
\\ [2 ex]
Institut f\"ur Theoretische Physik, Universit\"at W\"urzburg,\\
D-97074 W\"urzburg, Germany}
\end{center}
\vfill

\begin{center} \bfseries Abstract \end{center}

\noindent
We study the neutralino mass spectra and production cross sections at
an $e^+e^-$ linear collider
in superstring-inspired $\mathrm{E_6}$ models. These
models are characterized by the existence of new neutral gauge bosons
and singlet Higgs fields which also lead to an extended
neutralino sector compared to the Minimal
Supersymmetric Standard Model (MSSM) or the Next-To-Minimal
Supersymmetric Standard Model (NMSSM).
For different $\mathrm{E_6}$ models we develop scenarios with new light
exotic neutralinos which may offer important possibilities to
distinguish between the models.

\vfill
\newpage
\section{Introduction}

Superstring-inspired $\mathrm{E}_6$ models \cite{hewett} are
an interesting alternative for the phenomenology of 
low energy supersymmetry. They are characterized by the existence of
additional massive U(1) neutral gauge bosons and singlet superfields which
lead to an extended neutralino sector \cite{e6neutralinos}
compared to the Minimal 
Supersymmetric Standard Model (MSSM) \cite{MSSMneutralinos}
or the Next-To-Minimal
Supersymmetric Standard Model (NMSSM) 
\cite{NMSSMneutralinos}. Therefore the
investigation of the neutralino sector may offer an important
possibility to distinguish between the different models.

These models are based on an $\mathrm{E}_6$ gauge group, which is
broken to a low energy gauge group with one (rank-5 models) or two
(rank-6 models) new U(1) factors and therefore one or two new neutral
gauge bosons $Z'$ ($Z''$).
In the rank-5 models one or two additional Higgs singlet fields with 
nonzero vacuum expectation values (VEVs) appear whereas in the rank-6 case
two singlets are necessary to respect the experimental mass bounds for
new gauge bosons \cite{barger}. In all models we assume R-parity conservation,
that means that the lightest neutralino, the LSP, is stable and hence
invisible. 

The structure of the neutralino sector depends on the choice of the 
model generated by the respective mechanism of symmetry breaking
(for details we refer to \cite{later}):
\begin{itemize}
\item The \emph{rank-5 model with one singlet} contains 6 neutralinos which
are mixtures of photino, $Z$-ino, $Z'$-ino, 
two doublet higgsinos and
one singlet higgsino $\tilde{N}_1$. The $6\times 6$ neutralino mass
matrix depends
on six parameters: the $\mathrm{SU(2)_L}$, U(1)$_Y$ and
$\mathrm{U(1)'}$ gaugino 
mass parameters $M_2$, $M_1$ and $M'$, the ratio $\tan\beta = v_2/v_1$
of the VEVs of the
doublet Higgs bosons, the VEV $v_3$ of the singlet
Higgs and the trilinear Higgs coupling $\lambda$ in the superpotential.

\item An additional singlet higgsino $\tilde{N}_2$ appears in the
\emph{rank-5 model with two singlet fields}. 
Then the $7\times 7$ neutralino mixing matrix also
depends on the VEV $v_4$ of the second singlet Higgs.

\item The \emph{rank-6 model} with two extra U(1)-factors and therefore
two new gauge bosons contains 8
neutralinos (photino, $Z$-ino, $Z'$-ino, $Z''$-ino, two doublet
higgsinos and two singlet higgsinos $\tilde{N}_{1,2}$). 
Their masses and mixings are determined by 
8 parameters including 
the $\mathrm{U(1)''}$ gaugino mass parameter $M''$.
\end{itemize}

We analyze the neutralino mass spectra and
production cross sections at an $e^+e^-$ linear collider
in these $\mathrm{E_6}$ models
and inquire into the discrimination from MSSM and NMSSM. 
For this purpose
we work out scenarios with light singlet-like neutralinos
which are significantly different from the MSSM. 
Then we discuss the parameter regions where the cross
sections for neutralino pair production in the $\mathrm{E_6}$ models
are above an assumed discovery limit of 10 fb, so that   
a discrimination between the different supersymmetric models
may be possible. 

\newpage
\section{Scenarios and mass spectra}

Mainly from the experiments at the FermiLab collider there
exist lower mass bounds of about 600~GeV for new gauge bosons 
\cite{maeshima}. These limits imply
singlet VEVs of at least about 1500~GeV
in the rank-5 model with one singlet or in the rank-6 model. In
the rank-5 model with two singlets the VEVs can be smaller without
altering the phenomenology of the neutralino sector. 
In the following we choose $v_3 = (v_4 =)\;
1500$~GeV. Larger values lead to even larger masses of the $Z'$ bosons as well
as of the neutralinos with large singlet higgsino and $\tilde{Z}'$ components. 

All neutralino mass spectra and production cross sections are shown for the
whole range of the gaugino mass parameter 
$-1000$ GeV $\le M_2 \le +1000$ GeV. The value for 
$M_1$ is fixed by the usual gaugino mass relation
$M_1 = 5/3 \tan^2 \theta_W M_2 \approx 0.5 M_2$. 
For the additional gaugino parameters $M'$ and $M''$ we
partly assume the same unification relation but study also the impact of 
relaxing this condition.
For the parameters $\tan\beta$ and $\lambda$ which do not 
significantly influence the neutralino spectrum we use the exemplary values
$\tan\beta=2$ and $\lambda=0.14$. Identification of the MSSM parameter $\mu$ 
with $\lambda v_3$ implies a
value of $\mu = 210$~GeV for comparison of the $\mathrm{E}_6$ models
with the MSSM.

In all figures we show the experimentally excluded region from 
neutralino search at LEP \cite{lep}. 
Like in the NMSSM the existence of additional singlet-like
neutralinos does not substantially alter 
the respective MSSM parameter bounds
whereas very  
light singlet-like neutralinos cannot be excluded
\cite{NMSSMneutralinos}. 

In Fig.~1a the mass spectrum of the neutralinos is shown
in the rank-5 model with one singlet and $M'=M_1$. Here 
the four lighter neutralinos have mainly the same mixing 
character as in the MSSM for $|M_2| \lesssim 500$~GeV
whereas the neutralinos with large $\tilde{N}_1$ and $\tilde{Z}'$ components
lie on top of the spectrum (Fig.~1b, 1c)
and are kinematically excluded from production at the first stage of a
linear collider. 
Therefore discrimination of this model from the MSSM tends to be rather
difficult.

By relaxing the unification condition $M'=M_1$ and 
choosing a large constant value
$M'=10$~TeV a light neutralino appears in Fig.~2a
with mass $m_{\tilde{\chi}^0} \approx 42$~GeV nearly independent of
$M_2$. It is almost a pure singlet higgsino as can be seen from Fig.~2b.
The heaviest neutralino (the $\tilde{Z}'$, Fig.~2c)
has a mass of the order of $M'$ and is not shown in Fig.~2a.
Therefore the neutralino sector in this scenario is similar to the
NMSSM for suitable choice of the NMSSM parameters.

In the rank-5 model with two singlets always a very light 
singlet higgsino exists independently of the choice of the parameters
$v_3$, $v_4$, $\lambda$, $\tan\beta$ and 
even for $M'=M_1$ (Fig.~3a). 
In Fig.~3b the $\tilde{N}_1$ components of the neutralinos are
shown. Since the 
$\tilde{N}_2$ component is distributed among the neutralinos in the same
way the lightest neutralino with a mass of 
about 1~GeV is almost an equal mixture of the two singlet higgsinos. 
If additionally $M'$ is put to the large
value 10~TeV, two light singlet higgsinos appear (Figs.~4a, b).
In all cases the neutralinos with large $\tilde{Z}'$ and $\tilde{Z}''$
components are so
heavy that they probably lie outside the energy range of the
planned linear collider.

For the rank-6 model a similar characteristic
appears as for the rank-5 model with one singlet field.
With the assumption $M'=M''=M_1$ the four light neutralinos have 
mixing characters
as in the MSSM (similar to Fig.~1a) and 
the four new exotic neutralinos
are very heavy. If one of the new gaugino mass parameters ($M'$,
$M''$) has a large value (10~TeV), one light singlet higgsino appears
(analogue to Fig.~2a). 
If both parameters are fixed at large values there are two light exotic
neutralinos (similar to Fig.~4) which are mixtures of the two singlet
higgsinos.

\section{Cross sections}
For all the above described scenarios of the different $\mathrm{E_6}$ models
we have computed the cross sections for neutralino pair production and 
compared the results with the MSSM and NMSSM \cite{franke}. 
Here we use the rank-5 model with two singlets to present in Fig.~5 
the exemplary features of the production of light neutralinos as a
function of the c.m.\ energy of an $e^+e^-$ linear collider.
Fig.~5a shows the cross sections for $\tilde{\chi}^0_1\tilde{\chi}^0_2$-
and $\tilde{\chi}^0_2\tilde{\chi}^0_2$-production for $M_2=200$~GeV
and $M'=M_1$. Since the lightest neutralino is nearly a pure singlet higgsino
which does not couple to (s)fermions and gauge bosons, the 
$\tilde{\chi}^0_1\tilde{\chi}^0_2$-cross section reaches values above
a discovery limit of 10 fb only at the $Z'$ resonance.
This means that here a light singlet higgsino can be detected only in
connection 
with the identification of a new gauge boson $Z'$ which in our scenario has a
mass of 894~GeV. Then a discrimination between superstring-inspired
$\mathrm{E_6}$ models and the NMSSM is also possible.
Similar results are obtained for all $\mathrm{E_6}$ scenarios described in the
previous section with light singlet-like neutralinos.

The second lightest neutralino in this $\mathrm{E_6}$ model is of
nearly the same  
mixing type as the LSP in the MSSM. Therefore the $\tilde{\chi}^0_2$
pair production is almost identical with the production of two LSPs in the
MSSM. While the latter, however, remains invisible, the extended model
can be identified by the subsequent neutralino decays.
This is a typical example of an extended model where a neutralino, which 
would be 
invisible in the MSSM, is detectable because the mass spectrum
compared to the MSSM differs by the existence of
an additional light singlet higgsino.

In Fig.~5b we exploit a scenario with $M_2=-200$~GeV and
$M'=10$ TeV where the second and third neutralino are mixings of
all components but with large contributions
of about $25 \%$ of $\tilde{N}_1$ (see Fig.~4b) and $25 \%$ of $\tilde{N}_2$. 
Again one notes the 
$Z'$ resonance peak, but also outside of the resonance a nonminimal
neutralino is detectable by its direct production with a cross section
above 10 fb. In this case one could surely note a deviation from the
MSSM.
Also the discrimination from the NMSSM could be possible because
of the very light first neutralino which represents the
additional singlet eigenstate.
However, in a similar scenario in 
the rank-5 model with one singlet and large $M'$
a discrimination between the NMSSM and the $\mathrm{E_6}$ model seems only 
possible by also tracing signatures of an additional gauge boson. 

As already indicated, these results for a rank-5 model with two singlets
can be easily transferred to all other $\mathrm{E_6}$
models worked out in this contribution. 

\section{Conclusion}
We studied the neutralino sector of several superstring-inspired 
$\mathrm{E_6}$ models. It was shown that there exist several supersymmetric
scenarios with light singlet-like neutralinos compatible with the
current experimental constraints where a discrimination 
between $\mathrm{E_6}$ models, NMSSM and MSSM is possible.
Assuming grand unification of all gaugino mass parameters, however,
such scenarios are allowed only in a rank-5 model with two singlet fields.
Relaxing this condition one may find similar scenarios also for the
other $\mathrm{E_6}$ models.

The production of light neutralinos with significant singlet components
can reach cross sections large enough to allow detection and identification
of the respective model. Contrary, 
the cross sections for the direct production of nearly pure 
light singlet neutralinos 
is possible above a discovery limit of 10 fb only at a $Z'$ resonance in 
all studied $\mathrm{E_6}$ models. However, these models could be
identified by the  
decays of the heavier neutralinos with MSSM mixings into the light singlets.
Then the supersymmetric signatures sensitively depend on the dominant decay
channels which are to be studied in forthcoming works.

\newpage

\begin{figure}
\small
\setlength{\unitlength}{1cm}
\begin{picture}(16,17.4)

\put(1.1,17.1){(a) Mass spectrum}
\put(0.4,10.9){\includegraphics{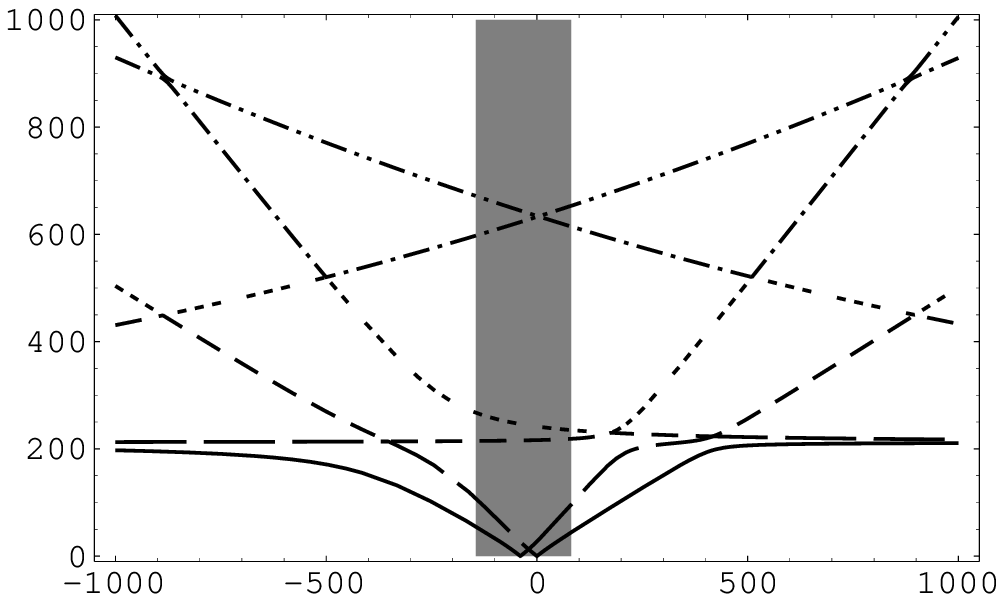}}
\put(0,14.1){\begin{sideways} $m_{\tilde{\chi}^0_i}/$GeV \end{sideways}}
\put(3.7,12.1){$M_2/$GeV}
\put(9.1,17.1){(a) Mass spectrum}
\put(8.4,10.9){\includegraphics{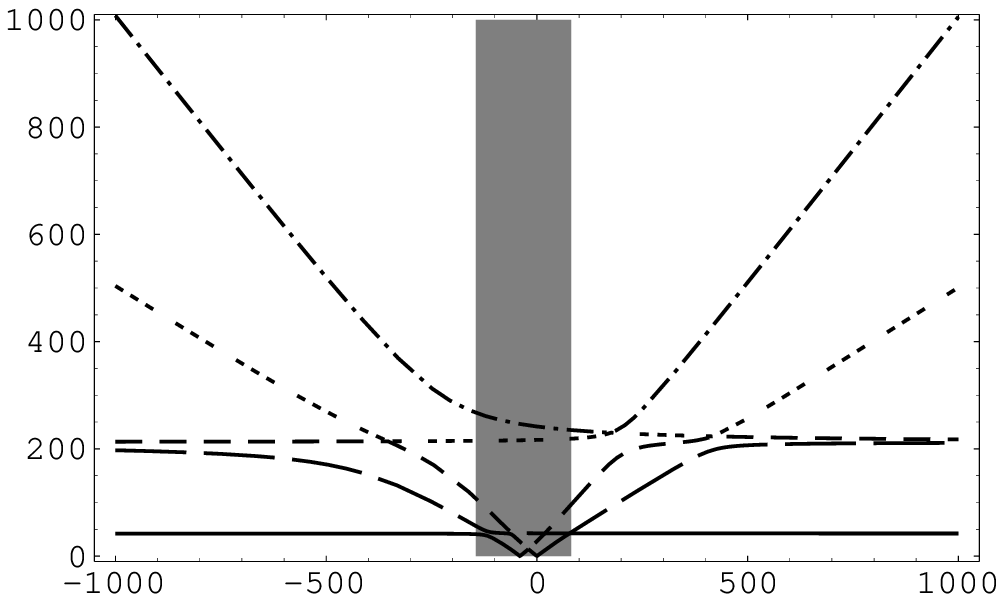}}
\put(8,14.1){\begin{sideways} $m_{\tilde{\chi}^0_i}/$GeV \end{sideways}}
\put(11.7,12.1){$M_2/$GeV}

\put(1.1,11.6){(b) $\tilde{N}_1$ components}
\put(0.4,5.3){\includegraphics{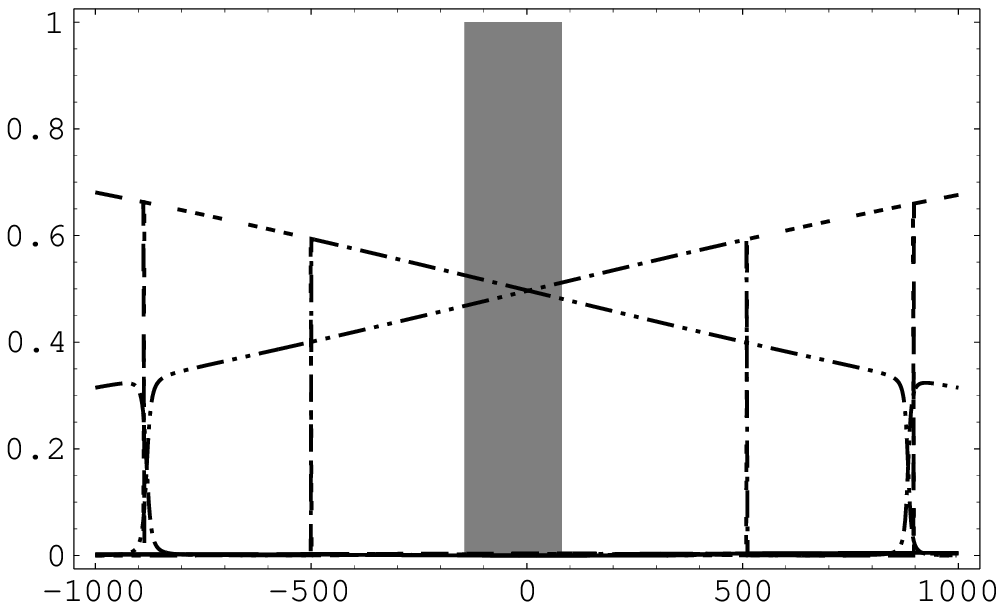}}
\put(-.05,8.5){\begin{sideways}
 $|\langle\tilde{N}_1|\tilde{\chi}^0_i\rangle|^2$ \end{sideways}}
\put(3.7,6.5){$M_2/$GeV}
\put(9.1,11.6){(b) $\tilde{N}_1$ components}
\put(8.4,5.3){\includegraphics{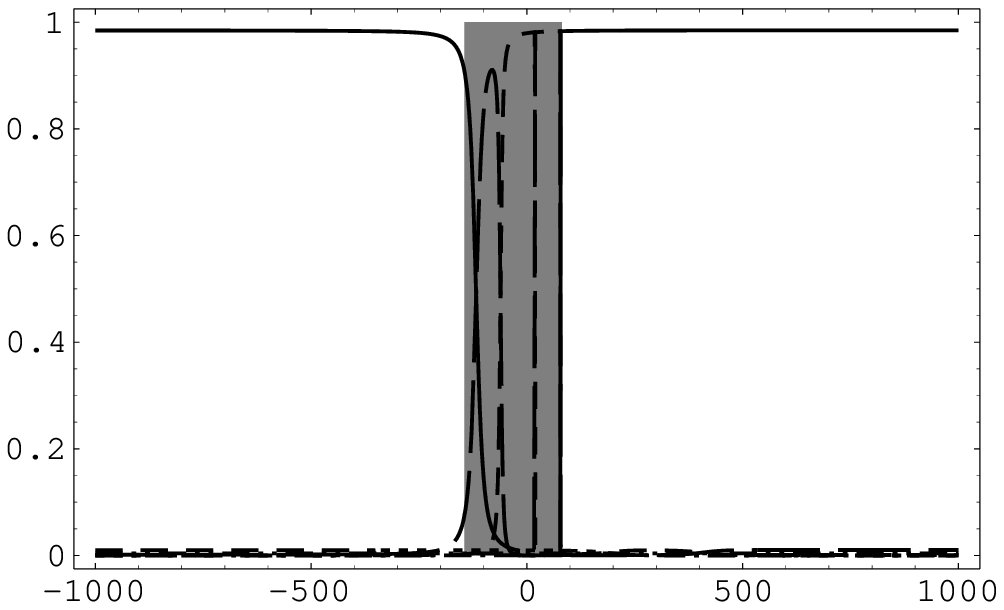}}
\put(7.95,8.5){\begin{sideways} 
 $|\langle\tilde{N}_1|\tilde{\chi}^0_i\rangle|^2$ \end{sideways}}
\put(11.7,6.5){$M_2/$GeV}

\put(1.1,6){(c) $\tilde{Z}'$ components}
\put(0.4,-0.3){\includegraphics{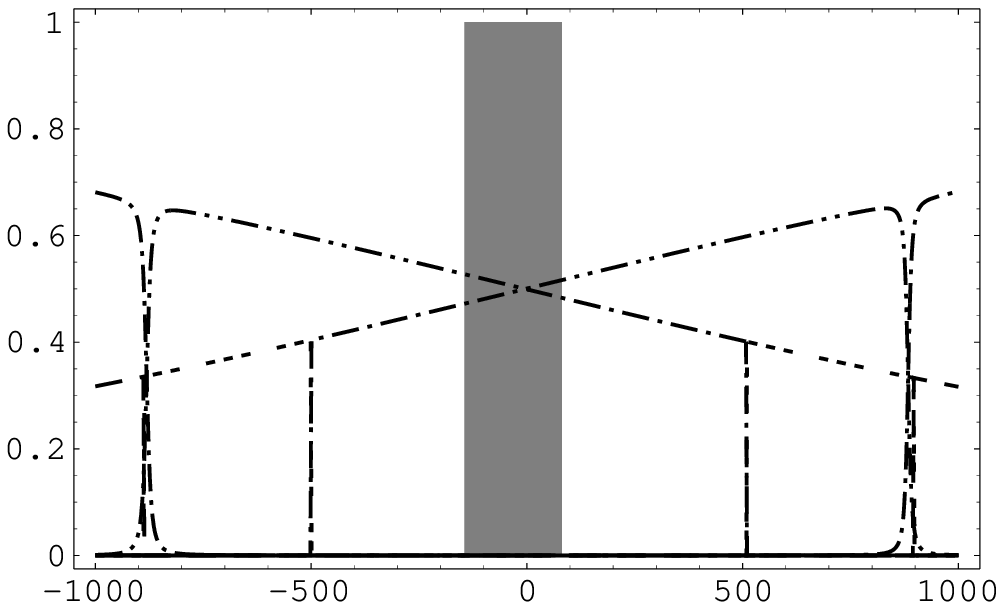}}
\put(-.05,2.9){\begin{sideways} 
 $|\langle\tilde{Z}'|\tilde{\chi}^0_i\rangle|^2$ \end{sideways}}
\put(3.7,.9){$M_2/$GeV}
\put(9.1,6){(c) $\tilde{Z}'$ components}
\put(8.4,-0.3){\includegraphics{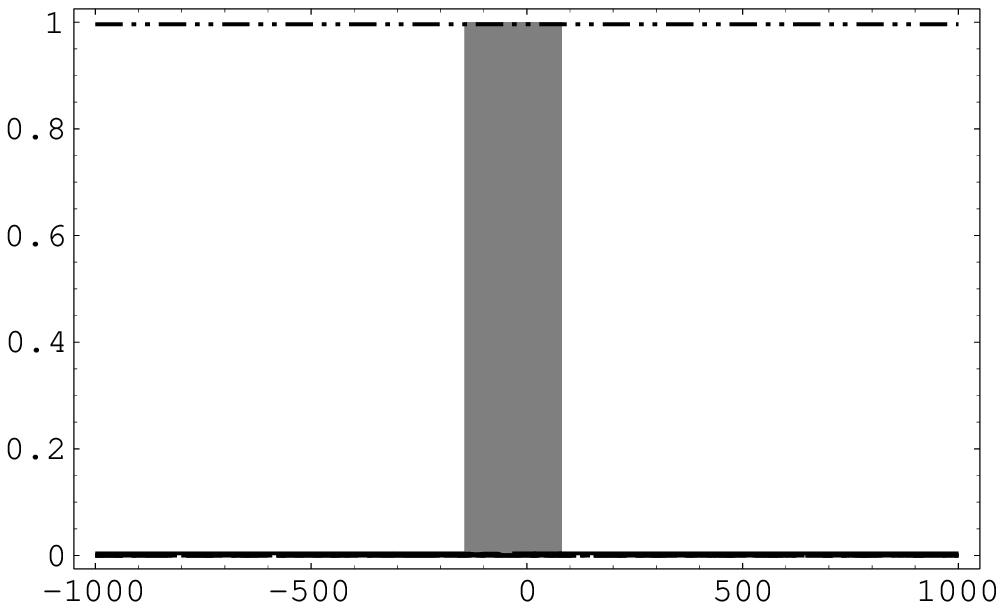}}
\put(7.95,2.9){\begin{sideways} 
 $|\langle\tilde{Z}'|\tilde{\chi}^0_i\rangle|^2$ \end{sideways}}
\put(11.7,.9){$M_2/$GeV}

\put(3.7,.2){Figure 1}
\put(11.7,.2){Figure 2}
\end{picture}

\normalsize\vspace{1mm}
Figure 1: Neutralinos in the rank-5 model with one singlet for
$M'=M_1$, 
$v_3 = 1500$~GeV, $\tan\beta=2$ and $\lambda=0.14$.
The shaded area marks the experimentally excluded $M_2$-region.

\vspace{2mm}
Figure 2: Neutralinos in the rank-5 model with one singlet for
$M'=10$ TeV,
$v_3 = 1500$~GeV, $\tan\beta=2$ and $\lambda=0.14$. The mass of the
$\tilde{\chi}^0_6$, of the order of $M'$, is not shown in (a).
The shaded area marks the experimentally excluded $M_2$-region.
\end{figure}

\clearpage

\begin{figure}[p]
\small
\setlength{\unitlength}{1cm}
\begin{picture}(16,11.8)

\put(1.1,11.5){(a) Mass spectrum}
\put(0.5,5.3){\includegraphics{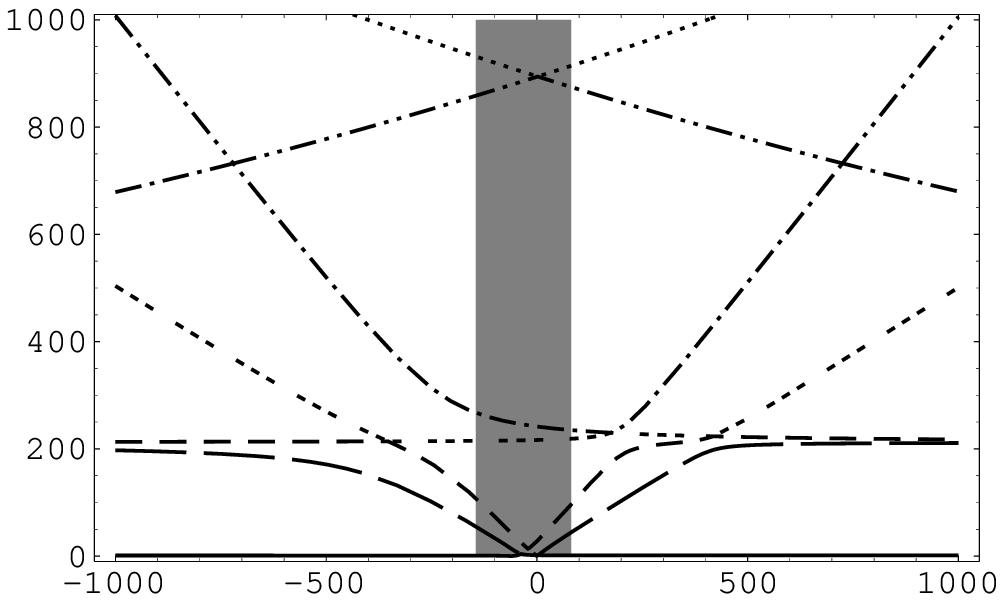}}
\put(.1,8.5){\begin{sideways} $m_{\tilde{\chi}^0_i}/$GeV \end{sideways}}
\put(3.7,6.5){$M_2/$GeV}
\put(9.1,11.5){(a) Mass spectrum}
\put(8.5,5.3){\includegraphics{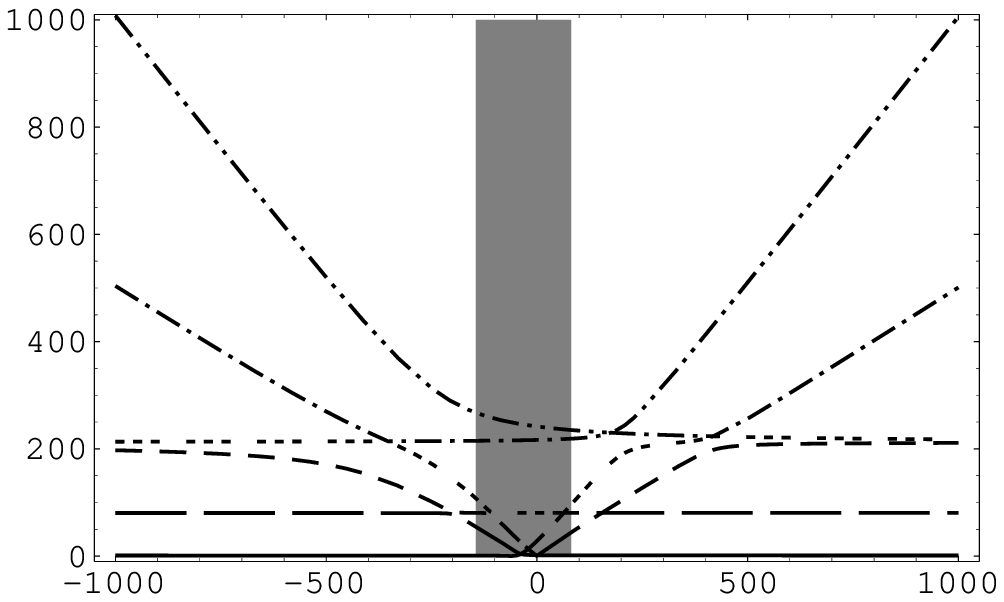}}
\put(8.1,8.5){\begin{sideways} $m_{\tilde{\chi}^0_i}/$GeV \end{sideways}}
\put(11.7,6.5){$M_2/$GeV}

\put(1.1,5.9){(b) $\tilde{N}_1$ components}
\put(0.5,-0.3){\includegraphics{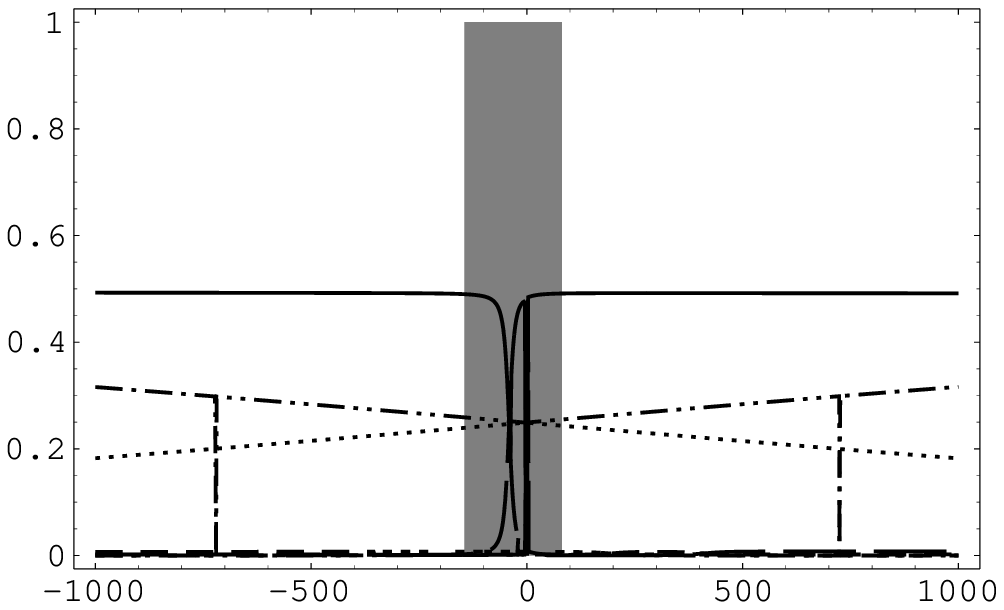}}
\put(0,2.9){\begin{sideways} 
 $|\langle\tilde{N}_1|\tilde{\chi}^0_i\rangle|^2$ \end{sideways}}
\put(3.7,.9){$M_2/$GeV}
\put(9.1,5.9){(b) $\tilde{N}_1$ components}
\put(8.5,-0.3){\includegraphics{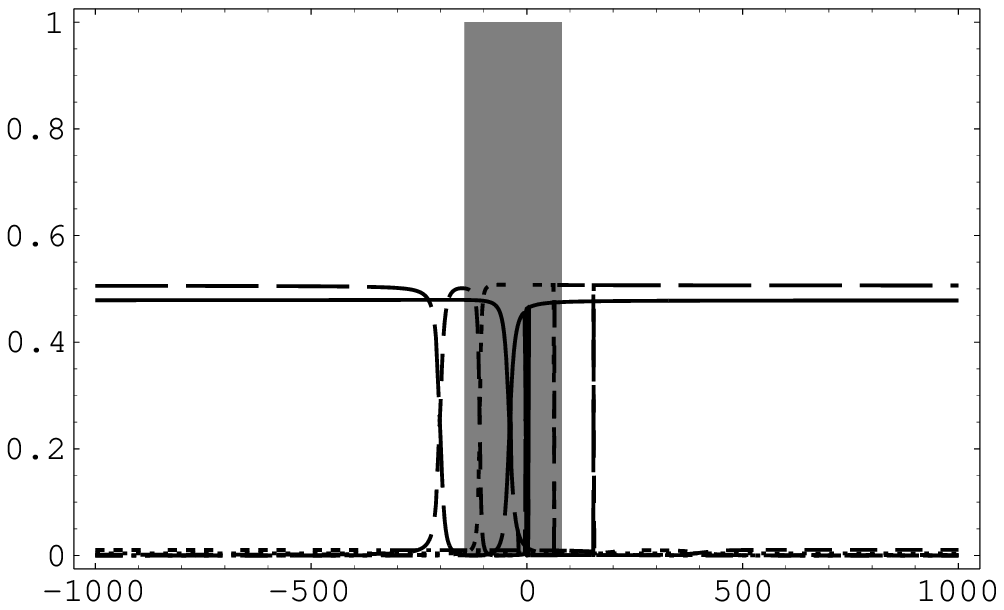}}
\put(8,2.9){\begin{sideways} 
 $|\langle\tilde{N}_1|\tilde{\chi}^0_i\rangle|^2$ \end{sideways}}
\put(11.7,.9){$M_2/$GeV}

\put(3.7,.2){Figure 3}
\put(11.7,.2){Figure 4}
\end{picture}

\normalsize\vspace{1mm}
Figure 3: Neutralinos in the rank-5 model with two singlets for
$M'=M_1$, 
$v_3 = v_4 = 1500$~GeV, $\tan\beta=2$ and $\lambda=0.14$.
The shaded area marks the experimentally excluded $M_2$-region.

\vspace{2mm}
Figure 4: Neutralinos in the rank-5 model with two singlets for
$M'=10$ TeV,
$v_3 = v_4 = 1500$~GeV, $\tan\beta=2$ and $\lambda=0.14$. The mass of the
$\tilde{\chi}^0_7$, of the order of $M'$, is not shown in (a).
The shaded area marks the experimentally excluded $M_2$-region.
\vspace{5mm}

\end{figure}

\begin{figure}[p]
\small
\setlength{\unitlength}{1cm}
\begin{picture}(16,4.9)

\put(1.8,4.9){(a) $M'=M_1$, $M_2=200$ GeV}
\put(0.5,-1.1){\includegraphics{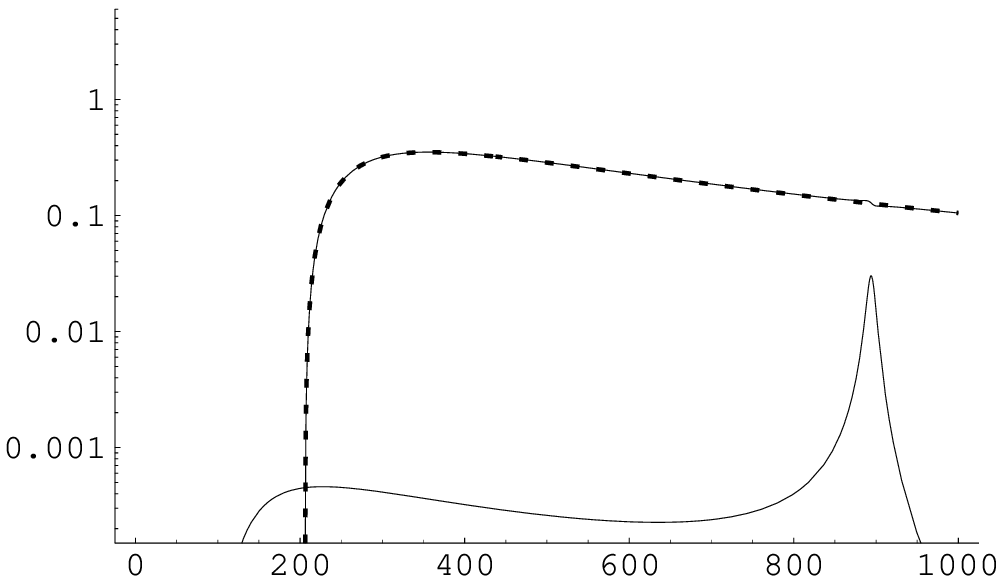}}
\put(.1,2.4){\begin{sideways} $\sigma/$pb \end{sideways}}
\put(4,.1){$\sqrt{s}/$GeV}
\put(2.5,4){\footnotesize 
 dashed: $\sigma_{\mathrm{MSSM}}(e^+e^-\to\tilde{\chi}^0_1 \tilde{\chi}^0_1)$}
\put(3.2,3.1){\footnotesize 
 $\sigma(e^+e^-\to\tilde{\chi}^0_2 \tilde{\chi}^0_2)$}
\put(3.8,1.4){\footnotesize 
 $\sigma(e^+e^-\to\tilde{\chi}^0_1 \tilde{\chi}^0_2)$}

\put(9.8,4.9){(b) $M'=10$ TeV, $M_2=-200$ GeV}
\put(8.5,-1.1){\includegraphics{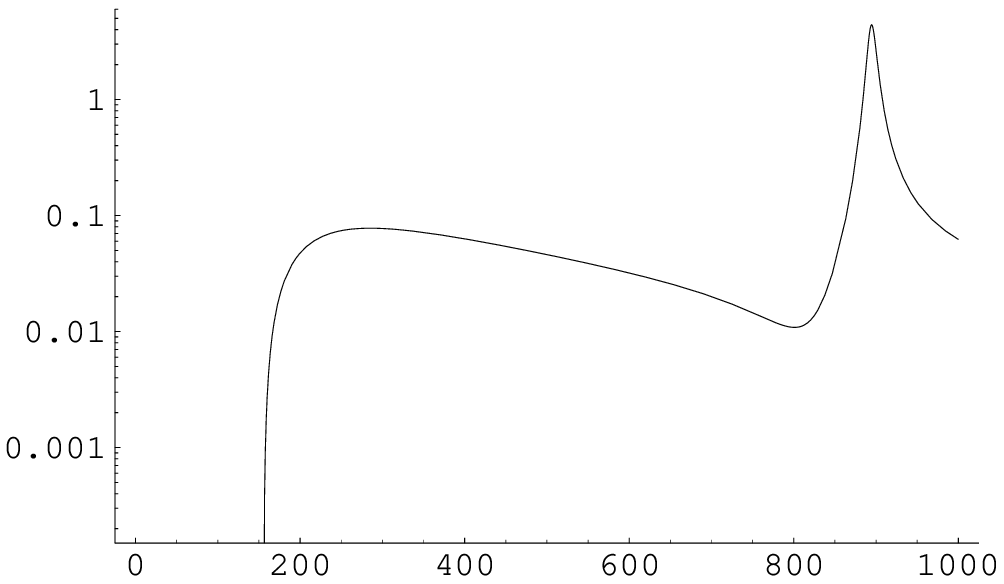}}
\put(8.1,2.4){\begin{sideways} $\sigma/$pb \end{sideways}}
\put(12,.1){$\sqrt{s}/$GeV}
\put(10.5,3.4){\footnotesize 
 $\sigma(e^+e^-\to\tilde{\chi}^0_2 \tilde{\chi}^0_2)$}
\end{picture}

\normalsize\vspace{1mm}
Figure 5: Total cross section for production of neutralinos in the
rank-5 model with two singlets for $v_3 = v_4 = 1500$~GeV,
$\tan\beta=2$, $\lambda=0.14$, $m_{\tilde{e}_L}=416$~GeV,
$m_{\tilde{e}_R}=120$~GeV, $m_{Z'}=894$~GeV and $\Gamma_{Z'}=12.5$~GeV.

\end{figure}

\end{document}